\documentclass[%
 reprint,
 amsmath,amssymb,
 aps,
]{revtex4-1}

\usepackage{graphicx}
\usepackage{dcolumn}
\usepackage{bm}

\usepackage[shortlabels]{enumitem}
\usepackage[caption=false]{subfig}
\graphicspath{{Figures/}}

\begin{document}


\title{Electrical Studies of Barkhausen Switching Noise in Ferroelectric \\lead zirconate titanate (PZT) and BaTiO$_3$:  Critical Exponents and Temperature-dependence}

\author{C. Flannigan}
\author{C. D. Tan}%


\author{J. F. Scott}
 \homepage{Corresponding author: jfs4@st-andrews.ac.uk}
\affiliation{
 Schools of Physics and of Chemistry, Univ. St. Andrews, St. Andrews KY16 9SS,  U. K.
}%



\begin{abstract}
Previous studies of Barkhausen noise in PZT have been limited to the energy spectrum (slew rate response voltages versus time), showing agreement with avalanche models; in barium titanate other exponents have been measured acoustically, but only at ambient temperatures. In the present study we report the Omori exponent (-0.95$\pm$0.03) for aftershocks in PZT and extend the barium titanate studies to a wider range of temperature.

\end{abstract}

\pacs{Valid PACS appear here}
\maketitle


\section{Introduction}\label{sec:level1}

Recent studies of crepitation (crackling noise) of ferroelectric lead zirconate titanate (PZT) and
barium titanate (BTO) samples during ferroelectric switching was limited to a single exponent
(energy versus time) for PZT and a single temperature (ambient) for BTO. In the present study,
using slew-rate data on voltage pulse output, we extend the PZT work to a wide temperature range
above and below room temperature, and measure additional exponents, including both amplitude
and aftershock (1894 Omori earthquake frequency) parameters. The data are compatible with avalanche 
models.\\

Ferroelectric and piezoelectric materials are used extensively in our day-to-day lives: in smart
phones alone there are piezoelectric microphones and speakers, not to mention the touch screen itself; and in MRI medical imaging equipment PZT is a usual detector element, limited by electrical noise. There are also ferroelectric memory systems (FRAM) present in smart cards as well as capacitors in every component of the electronics. The understanding and application of these materials are both at an advanced stage, with in-depth knowledge of its drawbacks being vital in selecting the right material for the given problem. There has been extensive research carried out into tackling some of the more obvious drawbacks such as fatigue, retention and electrical breakdown \cite{Scott1989FerroelectricMemories,Damjanovic1998FerroelectricCeramics,Dawber2005PhysicsOxides}. However, the noise and discontinuous nature of the ferroelectric response is still poorly understood. If the research were to continue on the same path, the small noise effects would become the limiting factors in the future, hence the importance of this investigation.\\

Additionally, some properties of noise in ferroelectrics have been shown to obey a power-law
and fit into a universality class \cite{Tan2019ElectricalDependence,Dahmen2009JerkyOf,Salje2017AnalysisDamping,Salje2019FerroelectricBaTiO3}. This is particularly interesting when applied to the study of
earthquakes. Earthquakes are not reproducible, but it has been found that the compression of porous
materials have the same universality class \cite{Baro2013StatisticalEarthquakes}, so it is possible to map the physics of earthquakes
onto this experiment in a way that is both easily reproducible and safe. Therefore, it is important to strengthen the link to earthquake statistics. The mechanism for polarisation in perovskite materials is well understood \cite{Scott1989FerroelectricMemories,Auciello1998TheMemories,Rabe2007ModernBackground,Ahn2004FerroelectricityHeterostructures,Schmidt1967PhaseFerroelectrics,Lines2001PrinciplesMaterials,Setter2006FerroelectricApplications}. There is a qualitative parameter referred to as the hardness of the ferroelectric/ferromagnetic response \cite{Auciello1998TheMemories,Setter2006FerroelectricApplications,Altpeter2016MaterialsMethods,Chilibon2012FerroelectricReview,Guyonnet2014FerroelectricWalls}. PZT shows a typical hard ferroelectric response (small coercive field; narrow hysteresis loop), and BTO shows a typical soft response.\\

There are several particular questions to address:  For example, in the simplest kind of avalanche (sand-piles) the particles are spherical and neither their shape nor mobility changes with temperature.  They typically exhibit single exponents over a wide range of measurements.  However, in a ferroelectric, the domains are more often planar; they greatly decrease in size and increase in mobility as phase transition temperatures are approached; and their anisotropy is more like rice grains than sand grains.  Does this affect their characteristic exponents?  Second, the Omori (1984) model of earthquakes gives an exponent for the time dependence of aftershocks, but not for their intensities.  Are these related?  The magnitude of shocks in our Barkhausen experiments is ca. a few attoJoules, roughly 30 orders of magnitude lower than in some earthquakes (Richter scale 5.0 = 1012 J).  It is rare in physics to find phenomena that scale over 30 orders of magnitude.\\

The maximum likelihood statistics used in this report can differentiate in a robust way between one exponent and a superposition of two.  In most cases we find two exponents describe the power spectra (based upon slew rates).  We tentatively interpret those as depinning from (1) point defects and (2) extended defects (such as threading dislocations, known to be in high concentration in PZT).\\

Barkhausen pulses have been studied previously in PZT and in other ferroelectrics and ferroelastics, particularly by three groups: Shur in Ekaterinburgh; Dul’kin and Roth in Israel; and Lupascu in Duisburg-Essen.  Table I lists some of the earlier studies, which generally did not emphasise power-law exponents.\\

\begin{table}[]
    \centering
    \begin{tabular}{l|c}
    \hline
        Material & References \\
         \hline
        LiTaO$_3$, LiNbO$_3$, Gd$_2$(MoO$_4$)$_3$  & a) b)\\
         PZT & b) c)\\
         PbTiO$_3$  & d) e) f)\\
         Ferroelastic Pb$_2$(PO$_4$)$_3$ & g)\\
        BaTiO$_3$   & h) k)\\
       Pb$_5$Ge$_3$O$_{11}$  & i) \\
        PbScTaO$_3$- and Bi$_{0.5}$Na$_{0.5}$TiO$_3$- mixed alloys & j) \\
        Pb(Mg$_{1/3}$Ta$_{2/3}$)O$_3$ , PbFe$_{1/2}$Nb$_{1/2}$O$_3$ & k)\\
        BST (Ba$_x$Sr$_{1-x}$TiO$_3$)   & l)\\
        \hline
             
    \end{tabular}
    \caption{Previous studies of Barkhausen pulses in ferroelectrics and ferroelastics\footnote{See also: D. Lupascu, Fatigue in Ferroelectric Ceramics and Related Issues (Springer, Berlin 2013)}}
    \label{tab:prev}
\end{table}

\begin{enumerate}[a)]
    \item V. Ya. Shur, E. L. Rumyantsev, D. V. Pelegov, V. L. Kozhevnikov, E. V. Nikolaeva, E. L. Shishkin, A. P. Chernykh \& R. K. Ivanov, Ferroelectrics 267, 347 (2010); this references also cites 9 related papers by Shur et al.; see especially I. S. Baturin, M. V. Konev, A. R. Akhmatkhanov, A. I. Lobov, \& V. Ya. Shur, Ferroelectrics 374, 136 (2008).
    \item Doru C. Lupascu , Jürgen Nuffer \& Jürgen Rödel, Ferroelectrics 290, 203 (2011); D. Lupascu, T. Utschig, V. Ya. Shur \& A. G. Shur, Ferroelectrics 290, 207 (2011). 
    \item  Jürgen Nuffer , Doru C. Lupascu \& Jürgen Rödel , Ferroelectrics 240, 1293 (2000).
    \item Hideo Iwasaki \& Mamoru Izumi, Ferroelectrics 37, 563 (2011).
    \item E. Dul'kin, J. Zhai, \& M. Roth, Phys Stat Sol A (2014).
    \item D. G. Choi \& S. K. Choi, J. Materials Science 32, 421 (1997). 
    \item E. K. H. Salje, E. Dul’kin, \& M. Roth,  Appl. Phys. Lett. 106, 152903 (2015). 
    \item A. G. Chynoweth, J. Appl. Phys. 30, 280 (1959); Phys. Rev. 110, 1316 (1958).
    \item I. J. Mohamad, L.Zammit Mangion, E. F. Lambson \& G. A. Saunders, J. Phys. Chem. Sol. 43, 749 (1982).
    \item E. Dul’kin, E. Mojaev, M. Roth, Wook Jo, \& T. Granzow, Scripta Mater. 60, 251 (2008); E. Dul’kin, B. Mihailova, M. M. Gospodinov \& M. Roth, J. Appl. Phys. 115, 084103 (2014).
    \item E. Dul’kin, A. Kania \& M. Roth, Phys. Stat. Sol. B1, 00145 (2016); Mater. Res. Express 1, 015105 (2014); E. Dul’kin \& M. Roth, J. Phys. Cond. Mat. 25, 155901 (2013)
    \item  E. Dul’kin, J. Zhai \& M. Roth, Phys. Stat. Sol. B252, 52111 (2015).
\end{enumerate}

\subsection{Pinning}
Henrick Barkhausen observed the discontinuous (jerky) nature in the ferromagnetic hysteresis loop in 1919 and used this to explain the presence of domains in the material \cite{Rudyak1971THEEFFECT}; this obscure behaviour was named after him. This mechanism that leads to the jerky nature of the response to the field. As a field is applied, a unit cell with a magnetisation (polarisation) opposing this field will be flipped when the field is larger than the coercive field, Bc (Ec) \cite{Rudyak1971THEEFFECT,Tebble1955TheEffect,Yamazaki2019ExperimentalWire}. This is most likely to occur just beyond the domain wall, as shielding effects are weakest here. This is equivalent to the perspective of the domain wall translating through the material \cite{Rudyak1971THEEFFECT,Tebble1955TheEffect,Yamazaki2019ExperimentalWire,Kagawa2016AthermalPoint,Salje2014CracklingMaterials}. As the domain wall approaches a defect, the translation is truncated at this point and the domain wall wraps around the defect – known as pinning. Barkhausen noise has been previously shown to obey a power-law probability distribution \cite{Tan2019ElectricalDependence,Dahmen2009JerkyOf,Salje2014CracklingMaterials},
which takes the form:
\begin{equation} \label{eq:power-law}
P[X]=\frac{\alpha-1}{X_{min}}\Big(\frac{X_i}{X_{min}}\Big)^{-\alpha}
\end{equation}
where $P[X]$ is the probability, $X_i$ is the parameter of the data which obeys a power-law, $X_{min}$ is the lower bound normalisation condition for the power-law – the value of $X_i$ that starts to obey the power-law \cite{Tan2019ElectricalDependence,Salje2014CracklingMaterials,Dahmen2009JerkyOf,Clauset2009Power-LawData}. To prevent divergence as $X\rightarrow0$, normalisation constants are introduced.\\

There has been a great deal of investigation into Barkhausen noise and comparing the crepitations involved to others that are statistically similar \cite{Kramer1996UniversalSheet,Sethna2001CracklingNoise,Salje2014CracklingMaterials,Dahmen2009JerkyOf,Dahmen1996HysteresisApproach,Dahmen2017MeanStatistics,Salje2019FerroelectricBaTiO3}. It is surprising how many different systems exhibit a crackling noise that obeys a power-law: crushing a chocolate bar wrapper in your hand \cite{Kramer1996UniversalSheet} and the famous ``Snap, Crackle, Pop!" when pouring milk into a bowl of Rice Krispies\texttrademark \hspace{0.225em}\cite{Sethna2001CracklingNoise,Salje2014CracklingMaterials} are some obscure examples. The aim of fitting to a power-law model is to extract the exponent. In a paper by Salje et. al (2019) \cite{Salje2019FerroelectricBaTiO3} several parameters of Barkhausen noise in BTO were fitted to a power-law. \\

The interesting thing here is the possibility of using this system to investigate earthquakes. It has been previously shown by Dahmen and Ben-Zion (2009) \cite{Dahmen2009JerkyOf} that Barkhausen noise and earthquake events share many statistical similarities. Similarly, Salje et. al (2019) fitted the acoustic data for their Barkhausen noise experiments on a BTO single crystal to Omori’s Law, with the exponent p=$1.0\pm0.2$ \cite{Salje2019FerroelectricBaTiO3}. \\

There have been many different mechanisms/theories proposed for the generation of aftershocks in earthquakes. Sholtz (1968) suggested this is due to the fractures or defects produced by residual stress left after a main fracture \cite{Scholz1968MicrofracturesSeismicity}. There are several other explanations follow a similar thought process of relaxation after stress \cite{Shaw1993GeneralizedDynamics,Ouillon2005Magnitude-dependentStudy,Wang2010PostseismicAftershocks}. In theories such as these, the reduction of aftershock rate is due to the reduction of probability of microfailure; this is due to the reduction of availability of microfailure sites \cite{Dyskin2019ResidualLaw}. This is an interesting perspective and is applicable to systems in this study; the stress on the material caused by switching of a 90$^{\circ}$ domain wall could then be relaxed by a subsequent domain switching, leading to an aftershock. Therefore, the diminishing rate is due to the diminishing availability of domains to be switched. \\

Omori’s law was developed in 1894 following analysis of a magnitude M=8 in Japan in 1891 \cite{Utsu1995TheActivity.,Guglielmi2017OmorisGeophysics}. It states that the frequency of aftershocks that follow the main event decreases with time according to the hyperbolic law:
\begin{equation}
n(t)=\frac{k}{c+t}
\end{equation}
where k and c are constants, and t is time \cite{Guglielmi2017OmorisGeophysics,Utsu1995TheActivity.,Dyskin2019ResidualLaw}. It is now more commonly replaced with a more general function \cite{Guglielmi2017OmorisGeophysics,Utsu1995TheActivity.,Dyskin2019ResidualLaw}:
\begin{equation}
n(t)=\frac{k}{(c+t)^{p}}
\end{equation}
where p is the power-law exponent and $0.7\leq p\leq 1.5$ \cite{Utsu1995TheActivity.,Dyskin2019ResidualLaw}. \\

The fitting of the Omori power-law to the jerk frequency data from ferroelectric switching experiments alludes to the possibility of designing reproducible experiments on earthquakes using these systems.

\subsection{Mean-field Theory}
The mean field model simplifies a system with complex behaviour by treating all contributions as a single averaged field. For many physicists, the mean field model is most likely encountered in the study of the many electron problem; where the repulsion on one electron from all the other electrons is approximated to a single potential \cite{meanfield}. Applying this to slip statistics, the microscopic details of the materials (locations and quantities of defects) are approximated to a mean field \cite{Dahmen2009MicromechanicalAvalanches,Dahmen2017MeanStatistics}. The model predicts the critical exponents of slip avalanches with a broad distribution of sizes (energy/amplitude of the jerks) \cite{Dahmen2017MeanStatistics}. There have been several notable experimental studies that have validated this model \cite{Tsekenis2013DeterminationPlasticity,Salje2014CracklingMaterials,Salje2019FerroelectricBaTiO3,Tan2019ElectricalDependence,Friedman2012StatisticsModel}. \\

The main assumptions of the model are the following:
\begin{itemize}
    \item The material has weak spots (eg. defects in the material and cracks in the rock for earthquakes) \cite{Dahmen2009MicromechanicalAvalanches,Dahmen2017MeanStatistics}.
    \item There is only one tuning parameter (the weakness of a weak site) \cite{Dahmen2009MicromechanicalAvalanches}.
\end{itemize}

As shear stress increases, the weak spots are stuck until the local shear stress exceeds the random threshold of failure for that site \cite{Dahmen2009MicromechanicalAvalanches,Dahmen2017MeanStatistics}; this leads to a slip by a random distance. The release in stress from the failure of a weak site is redistributed to other weak points, with some probability of this breaching the failure threshold of the sites that now have an increased shear stress. In this study, the applied field will cause the domain wall to move and this will be the mechanism for shear stress on the weak sites, and the slip distance will be observed as the jerk in current. \\

The probability distribution of energies of jerks, P(J), is proportional to the energy of the jerk to some critical exponent, $\epsilon$:
\begin{equation}
P(J)\propto J^{-\epsilon}
\end{equation}
where energy scales as:
\begin{equation}
J \sim \int v(t)^2 dt
\end{equation}
v(t) is proportional to the instantaneous growth rate of the avalanche \cite{Dahmen2009MicromechanicalAvalanches,Dahmen2017MeanStatistics}.\\

The probability distribution is also a function of shear stress. In the experiments in this study, the avalanche energies are taken over the entire stress range. Therefore, it is important to integrate over stress to predict the exponent correctly. This leads to a larger exponent $\epsilon=-\frac{5}{3}$ compared with $\epsilon=-\frac{4}{3}$ when not stress integrated \cite{Dahmen2009MicromechanicalAvalanches,Dahmen2017MeanStatistics,Tsekenis2013DeterminationPlasticity,Salje2014CracklingMaterials,Salje2019FerroelectricBaTiO3,Tan2019ElectricalDependence,Friedman2012StatisticsModel}.

\section{Experimental}

\subsection{Samples}

Earlier studies of Barkhausen noise in ferroelectrics \cite{Salje2019FerroelectricBaTiO3,Mai2015TheFilms,Salje2015AcousticPb3PO42,Lupascu2003TheMeasurements} emphasised a few well-known materials; a full listing is given in Table \ref{tab:prev}.\\

The specimens used in the present experiments were described in earlier papers. \cite{Tan2019ElectricalDependence,Salje2019FerroelectricBaTiO3} Three ceramic PZT specimens were examined, emphasising  a commercial lead zirconate titanate (referred to as PZT throughout) from PI Ceramic Lederhose, Germany \cite{PiezoceramicMaterials}, labelled PIC255 by the manufacturer, and one single-crystal barium titanate sample.   In addition BaTiO$_3$ ceramics were prepared that were Fe-doped.

\subsection{Electronics}
A high voltage amplifier was used to generate pulses up to 1.5 kV in the project but can generate a potential difference of up to 40 kV. The FE-module allows for hysteresis loops to be measured \cite{TFManual}; this can be replaced by other modules to probe other electronic behaviours of materials \cite{TFManual}. For the high temperature measurements, the sample was placed in a aixACCT piezo sample holder which could be used to set the temperature up to 473 K. For cold temperature measurements, the sample was placed in a closed-cycle cryostat, which could cool down to approximately 80 K by means of the Gifford-McMahon refrigeration cycle \cite{Gifford1966TheCycle}.  \\

\subsection{Data}
The triangular waveforms generated using the TF-Analyzer software were applied to the sample via the probes and the current and polarisation responses were measured. These waveforms were manually generated using the software’s manual waveform generator. The maximum number of points that can be measured using the software is 1000, hence it was important to choose the recording region to maximise the number of useful data points. The coercive field of ferroelectrics is frequency dependent \cite{Scott1996ModelsFilms,Viehland2000Random-fieldReversal}; the voltage ramp-rate was kept constant throughout the experiments to ensure this was not an issue.

\section{Avalanche Analysis of Exponents}

\subsection{Power-law Histogram (PLH)}
This is a straightforward method for extracting the exponent. The natural logarithm of the data is taken and the data is divided into bins of equal logarithmic width using the histcounts function in Matlab. This is preferred to binning linearly then taking the natural logarithm due to compilation errors in the code when taking the natural logarithm of an unpopulated bin. The probability was scaled with the width of the bins using equation \ref{eq:LogScalePopulation} so that the weighting was correlated to the width. \\
\begin{equation} \label{eq:LogScalePopulation}
    P[X_i] = \frac{\Tilde{P}[X_i]}{\exp{(X_i)}-\exp{(X_{i+1})}}
\end{equation} 
where $X_i$ is the position at the start of the bin (hence $X_{i+1}$ is the position at the end of the bin), $\Tilde{P}(X)$ is the raw population and P(X) is the re-scaled population. The data is denoted X here and elsewhere to acknowledge that this can be several parameters of the system. It is important to re-scale the population, as without this, the linear fit produces an exponent of $\eta+1$ instead of $\eta$ due to an increase in bin width with increasing X. The distribution was normalised using the trapz function in Matlab; this was generally poor as it uses the trapezoidal method \cite{TrapezoidalKingdom}. However, this was acceptable in this study as the focus is on the shape of the distribution produced, not the specific values.

\subsection{Maximum-Likelihood (ML)}
After initial shocks, whether voltage pulses in ferroelectrics or real earthquakes, the frequency of aftershocks subsides.  This decrease can be fitted in two ways:  (1) It can simply be fitted via least squares to an exponential.  This gives a plausible result but it suffers from the fact that it is not obvious where to begin and end the data set being fitted.  A more rigorous approach is to use the maximum likelihood procedure, which requires that there be a plateau in certain graphs of the raw data.  This procedure is reviewed in refs \cite{Tan2019ElectricalDependence,Salje2019FerroelectricBaTiO3,Clauset2009Power-LawData,Rice2007MathematicalAnalysis,Vassiliev2017SamplingTechniques}.  It is useful in picking out unambiguously where the data do not satisfy a single power law.  In such cases the data may be a superposition of two or more power laws, or they may not be power laws at all (e.g., a logarithmic or algebraic decay).  In the present study the “waiting time” data between aftershocks do not satisfy this power law requirement.

\subsection{Resulting Omori Exponent Aftershocks in PZT}
\begin{figure}
    \centering
    \includegraphics[width=8.6cm]{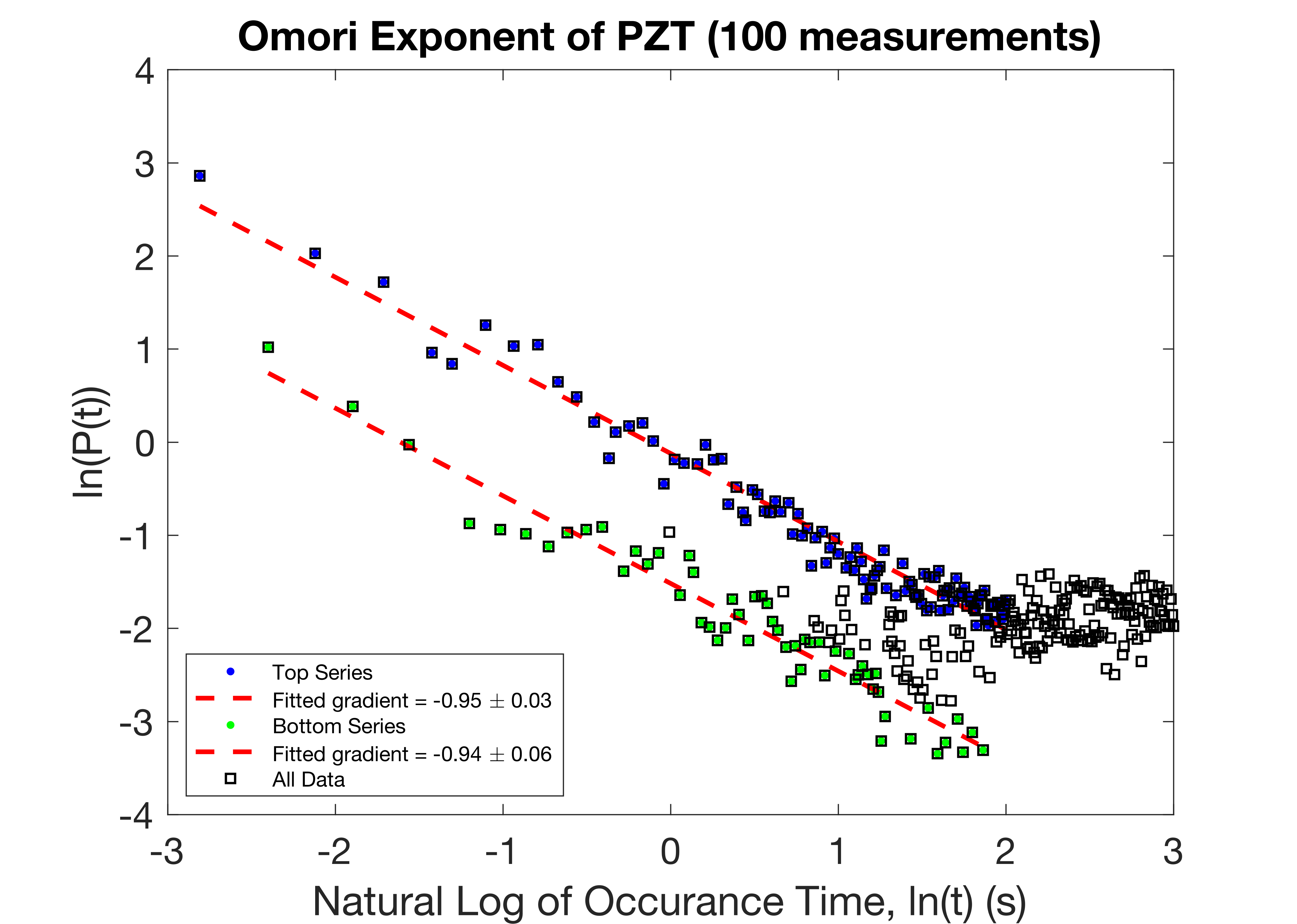}
    \caption{Power-law histogram of event occurrence time for PZT at room temperature for 100 measurements. Event times are taken relative to the time of the first event. The Omori exponent, p, is extracted from the gradient of the fit. The two series are still seen, but the gradients are now far more similar. Note: the trapezoidal normalisation is very poor in this case, leading to several data points having greater than 100\% probability. Only the gradient of the line is important so this will not affect the conclusions.}
    \label{fig:omori_100}
\end{figure}

Data for aftershocks in PZT: The Omori Law.  Fig. \ref{fig:omori_100} shows the results of 100 runs on PZT at room temperature (298K).  The upper curve is for large amplitude voltage pulses (ca. 500-1000 attoJ) and gives an Omori exponent of p = 0.95$\pm$0.03.  The traditional value in earthquakes since the 8.0-magnitude Japanese quake of 1894 is 1.0; and in barium titanate at room temperature Salje et al. \cite{Salje2019FerroelectricBaTiO3} found 1.0$\pm$0.2.  We emphasise that \cite{Salje2019FerroelectricBaTiO3} measured shocks acoustically whereas we measure them via voltage output; the two signals are not a priori identical (especially in the case of 90-degree domains in pure ferroelastics such as lead phosphate which are not ferroelectric).

\subsection{Amplitude Exponent in PZT}
\begin{figure}[h!]
\subfloat[\centering Amplitude Power-law Histogram PZT at RT]{%
  \includegraphics[width=8.6cm]{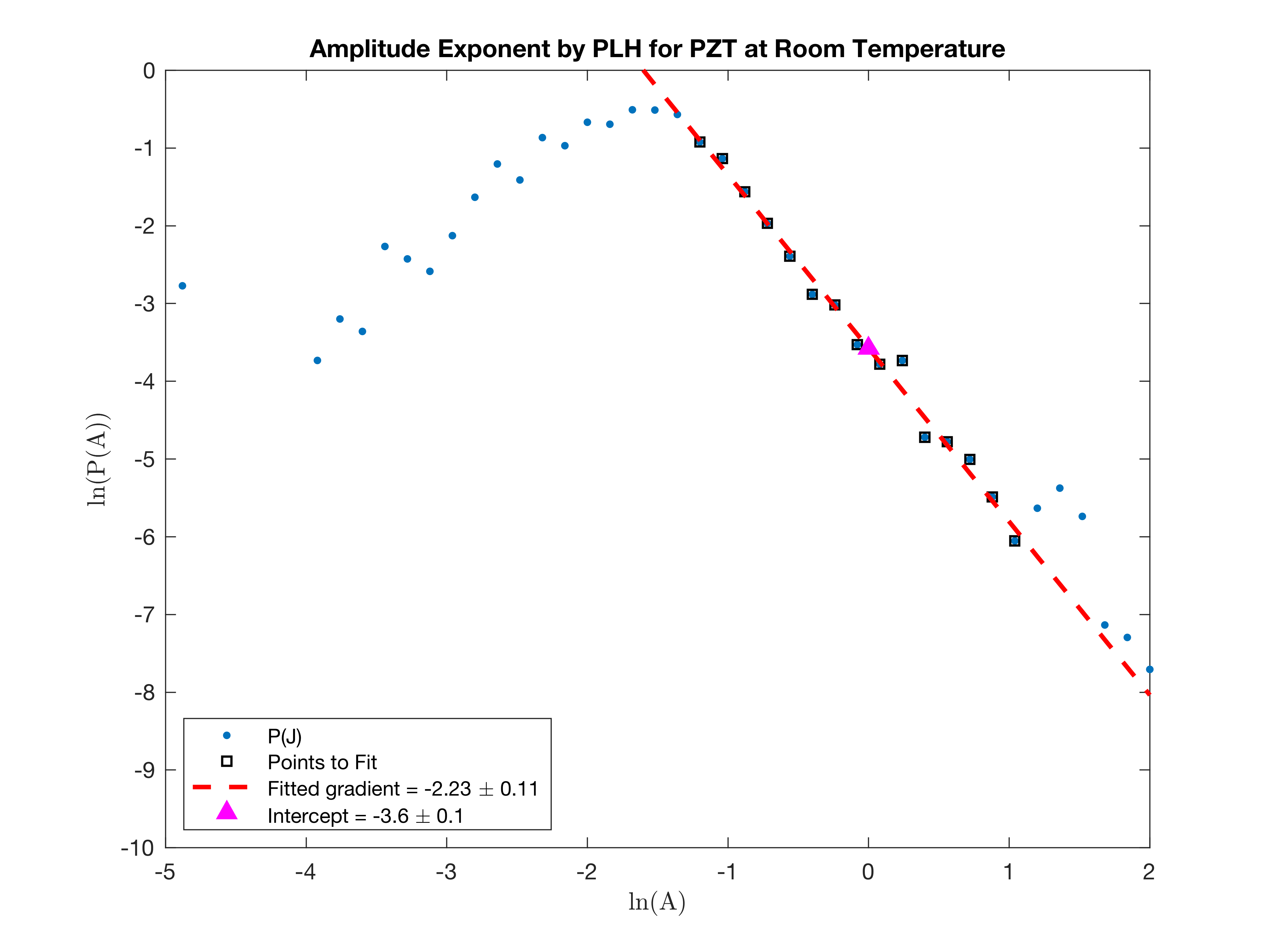}%
}

\subfloat[Amplitude Maximum-likelihood PZT at RT\label{fig:amp_rt_ml}]{%
  \includegraphics[width=8.6cm]{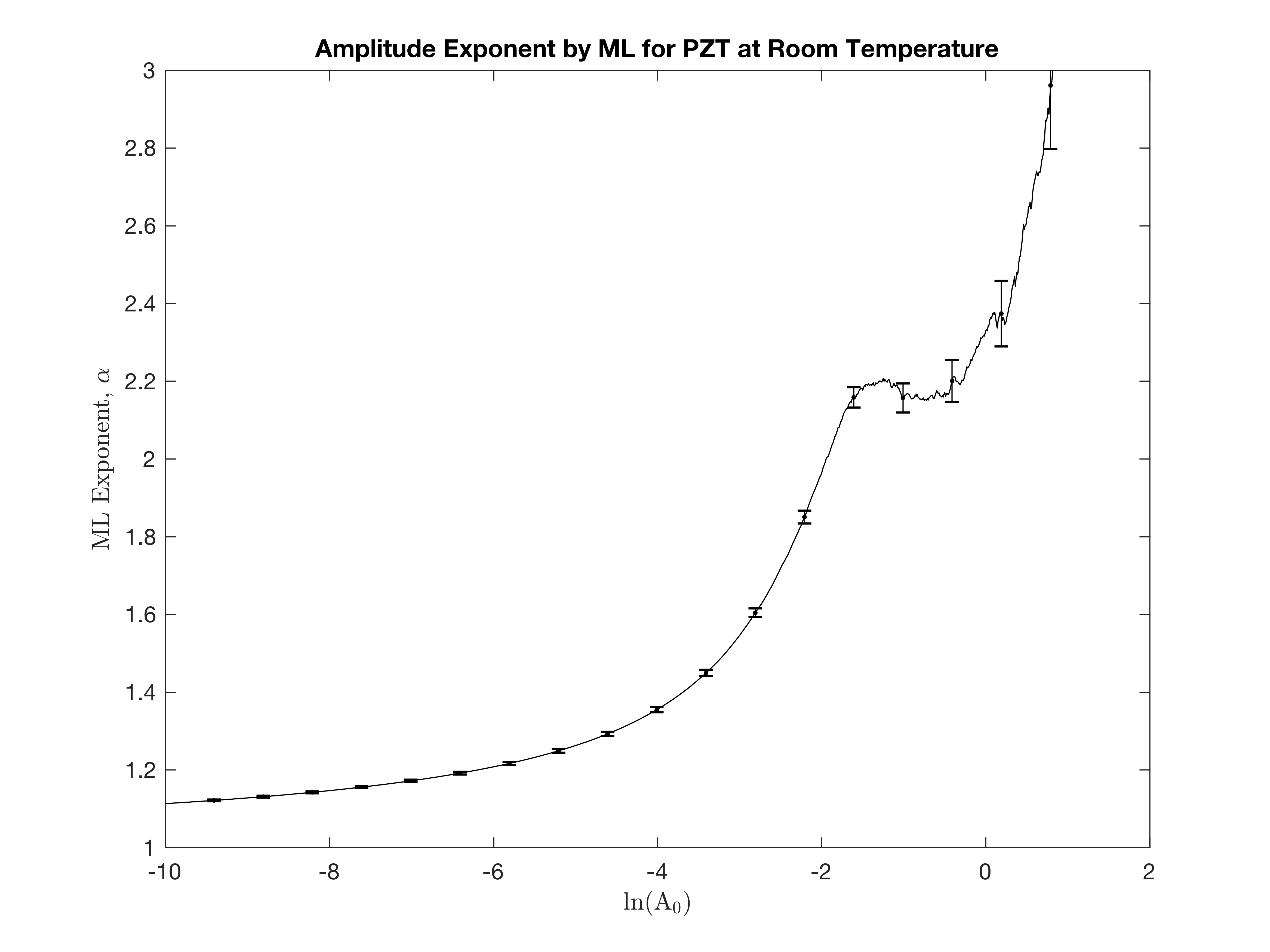}%
}
\caption{\textbf{(a)} Power-law histogram and \textbf{(b)} maximum-likelihood plot for amplitude exponent of PZT sample at room temperature. Consistent evidence between the two analysis methods suggesting the exponent $\alpha=2.23\pm0.11$. The region of the plateau in the ML plot and the region of straight line in the PLH are much shorter than that for the energy exponent in Figure \ref{fig:energy_rt}. The difference between amplitude data and energy data is that the former pick out the maximum amplitude in a cluster of voltage pulses; however, if these neighbouring pulses are not resolved, there is no new information in them, and the two exponents are not independent.}
\label{fig:amp_rt}
\end{figure}

Fig. \ref{fig:amp_rt} illustrates the fitting of data to the amplitude exponent, $\alpha$ in PZT at room temperature.  
Here the amplitude is the number of aftershock pulses within a certain bin size.  Our value is 2.25$\pm$0.11, in good agreement with the value of 2.23 for barium titanate.\cite{Salje2019FerroelectricBaTiO3}

\subsection{Duration Exponent in PZT}

\begin{figure}[h!]
\subfloat[\centering Duration Power-law Histogram PZT at RT]{%
  \includegraphics[width=8.6cm]{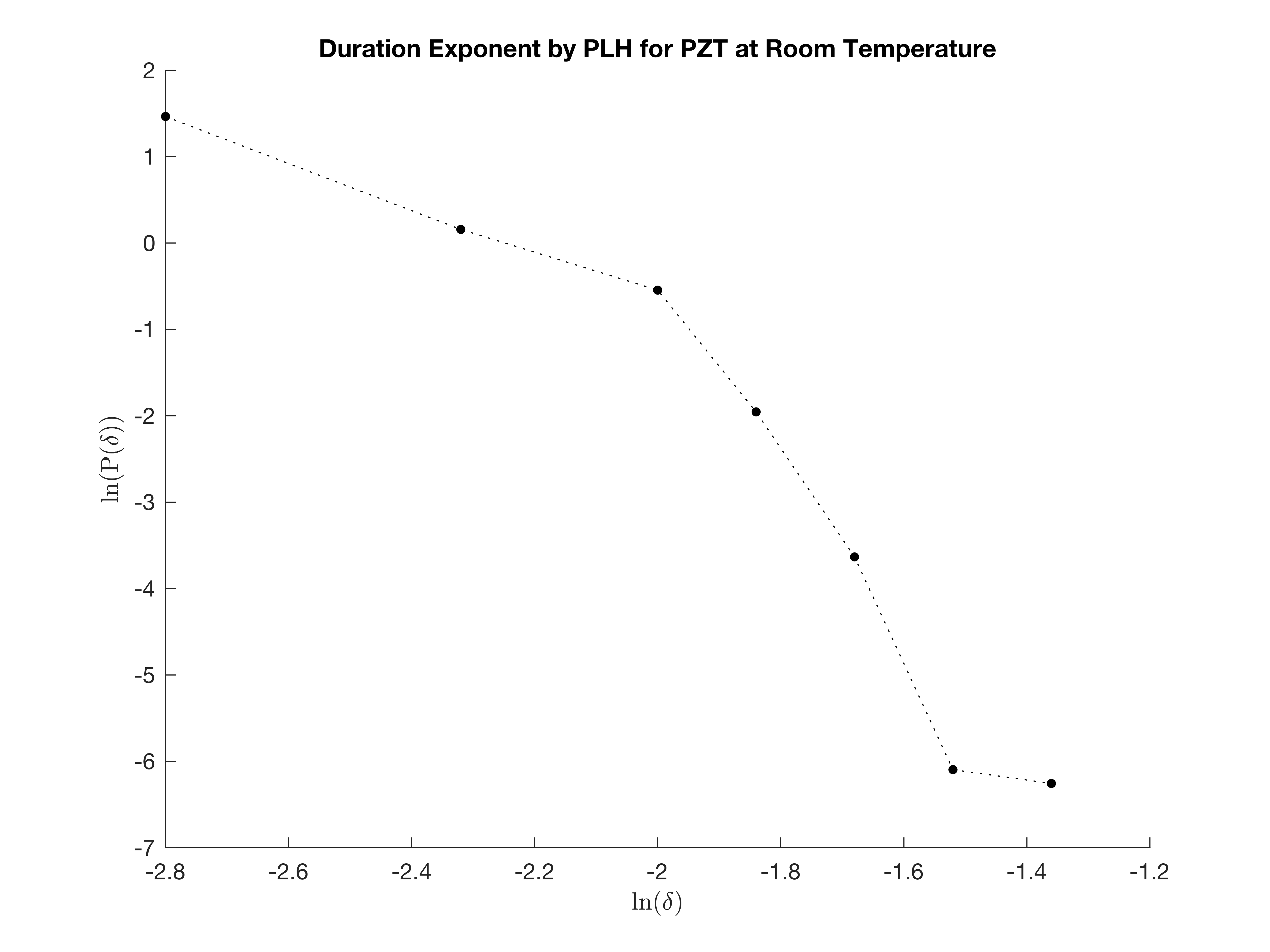}%
}

\subfloat[Duration Maximum-likelihood PZT at RT\label{fig:amp_rt_ml}]{%
  \includegraphics[width=8.6cm]{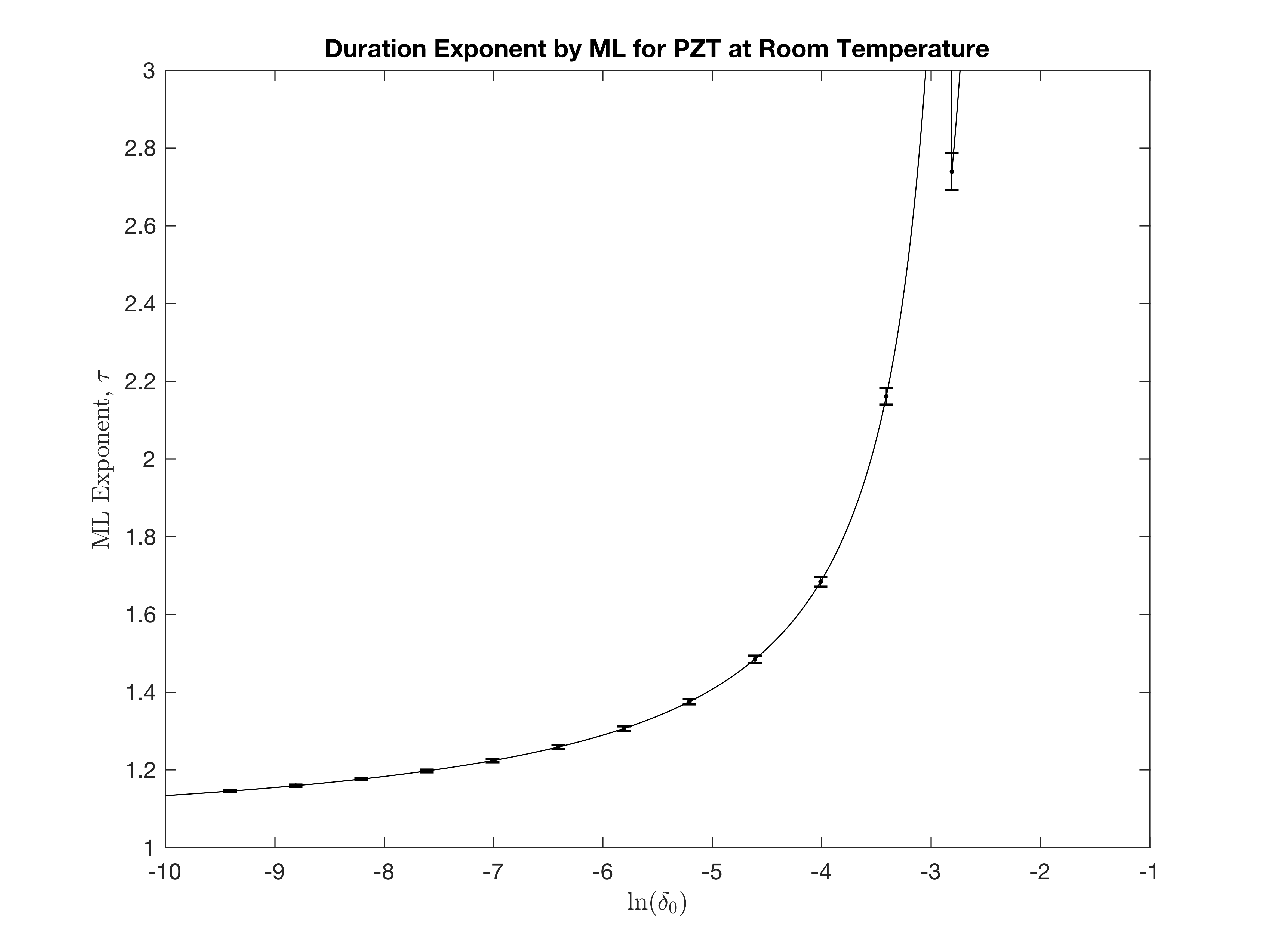}%
}
\caption{\textbf{(a)} Power-law histogram and \textbf{(b)} maximum-likelihood plot for duration exponent of PZT sample at room temperature. There is no straight line in the PLH and no plateau in the ML curve, hence this distribution does not obey a power-law. This is likely due to the sampling rate giving poor time resolution.}
\label{fig:dur_rt}
\end{figure}

Fig. \ref{fig:dur_rt} illustrates the duration of aftershocks for PZT at ambient temperature.  The data do not exhibit a plateau in the maximum likelihood analysis, and so we conclude that the present data are insufficient to imply a power law dependence or to yield an unambiguous exponent.

\subsection{Further Data Versus Temperature for Barkhausen Exponents in PZT}

Although we have presented some data \cite{Tan2019ElectricalDependence} on energy exponents in PZT previously, we extend those here in Table II and Figs. 4 and 5.  These show three sets of exponents observed: One near 1.4; one near 1.7, and a third near or above 2.0.  In Ref. \cite{Salje2019FerroelectricBaTiO3} three regimes were also observed for BaTiO$_3$, but the shorter middle regime was not attributed to a third, distinct exponent, but instead to a transition region with Poisson statistics. We can only speculate about their physical origins (depinning from point or extended defects).

\begin{figure}[h!]
    \centering
    \includegraphics[width=8.6cm]{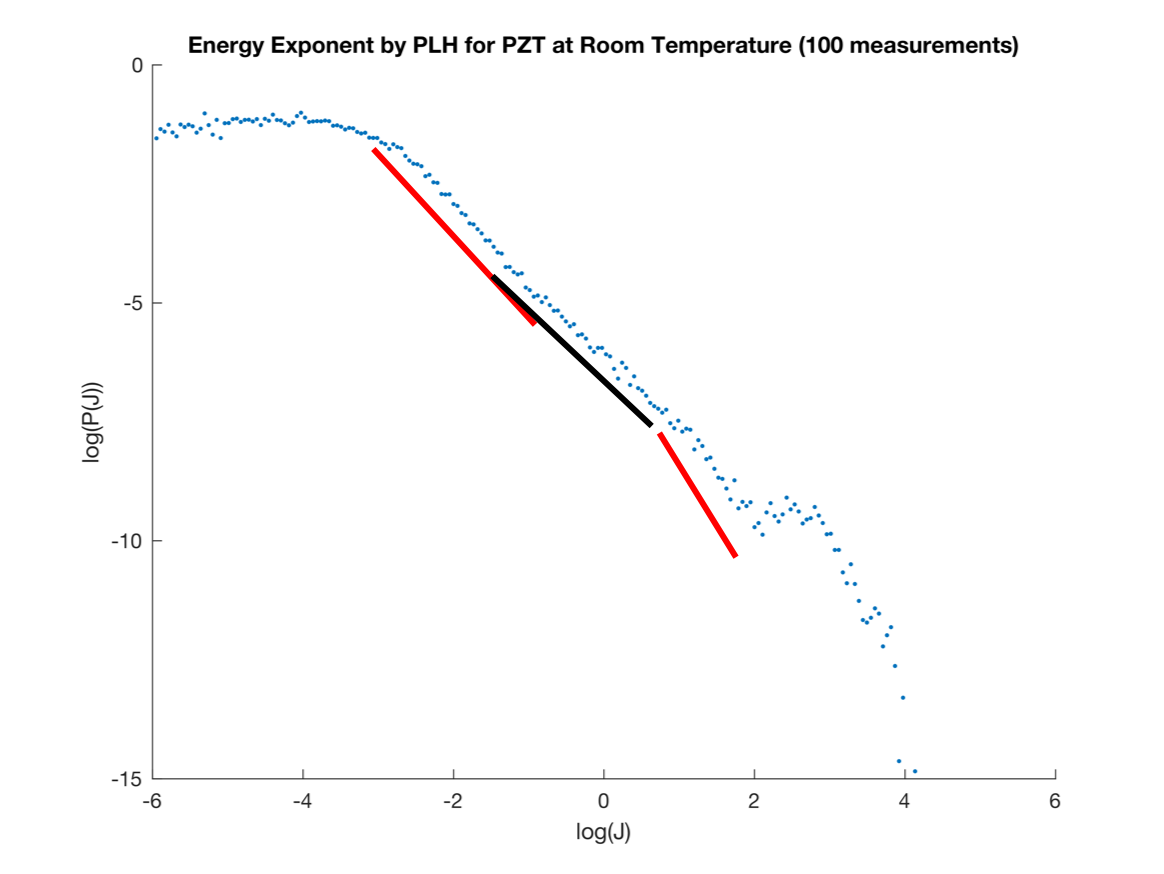}
    \caption{Power-law histogram for PZT sample at room temperature with 100 measurements. There is a mixture of three exponents with the values summarised in Table \ref{tab:energy_rt_100}.}
    \label{fig:energy_rt_100}
\end{figure}

\begin{table}[h]
    \centering
    \begin{tabular}{|c|c|}
    \hline
        Exponent & Value \\
        \hline
         $\epsilon_1$ & $1.47\pm0.04$\\
          $\epsilon_2$ & $1.79\pm0.08$\\
           $\epsilon_3$ & $2.04\pm0.19$\\
           \hline
    \end{tabular}
    \caption{Summary of linear fitting to the three exponents seen in the PLH for PZT at room temperature with 100 measurements (Figure \ref{fig:energy_rt_100})}
    \label{tab:energy_rt_100}
\end{table}

\onecolumngrid

\begin{figure}[h!]
\subfloat[Energy Power-law Histogram PZT at 259 K]{%
  \includegraphics[width=8.6cm]{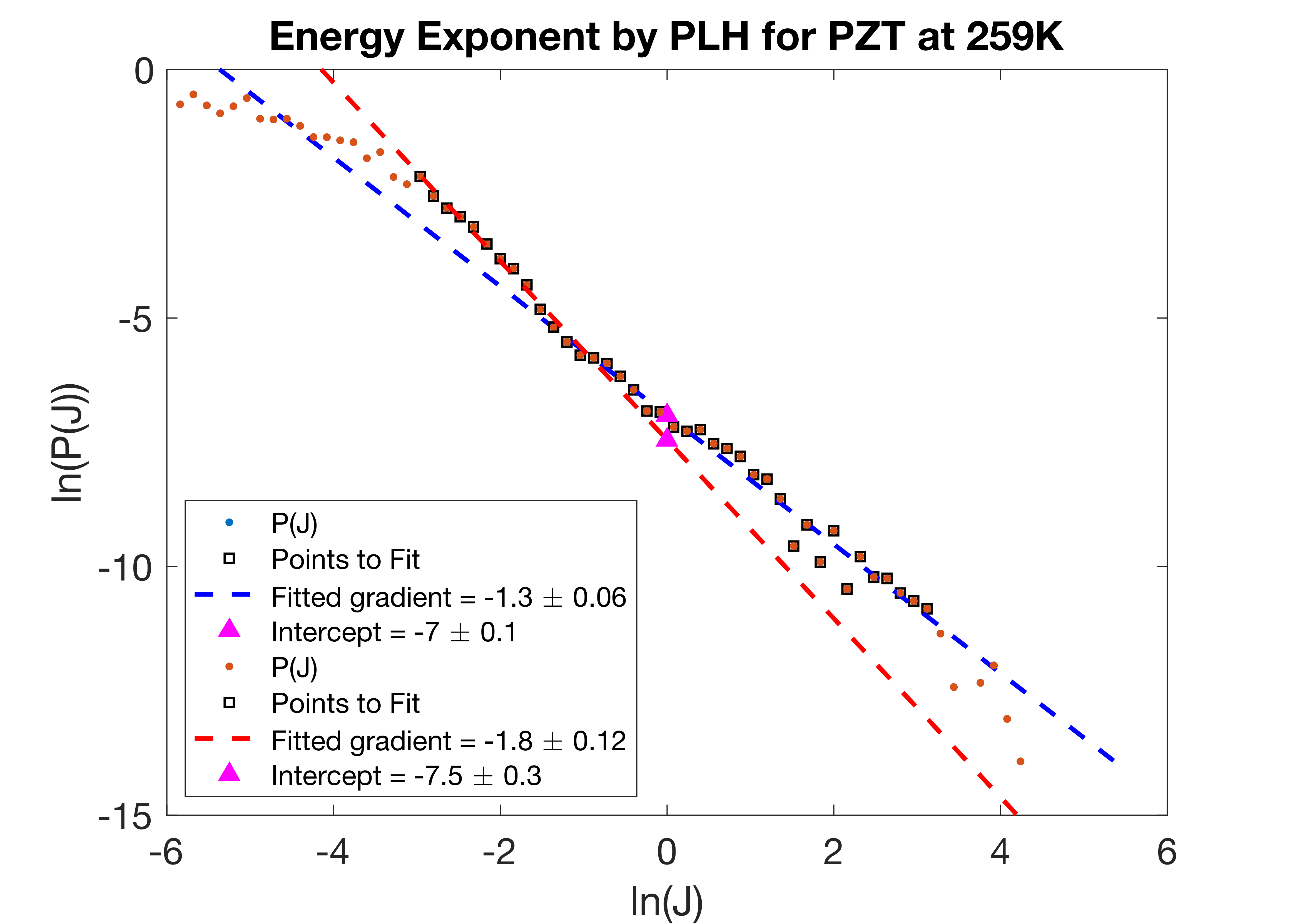}%
}
\subfloat[Energy Maximum-likelihood PZT at 259 K\label{fig:259_ml}]{%
  \includegraphics[width=8.6cm]{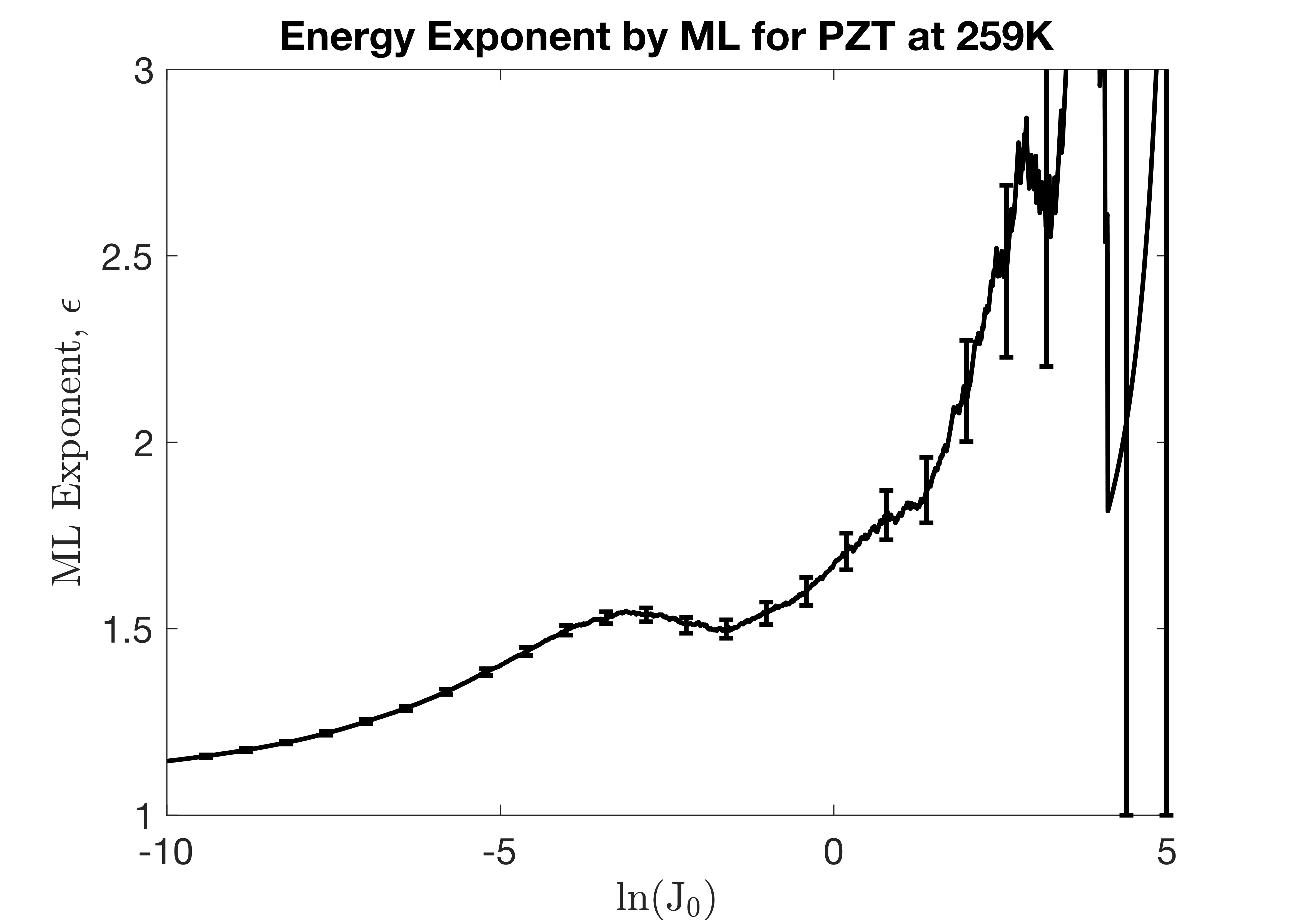}%
}\\
\subfloat[Energy Power-law Histogram PZT at 413 K]{%
  \includegraphics[width=8.6cm]{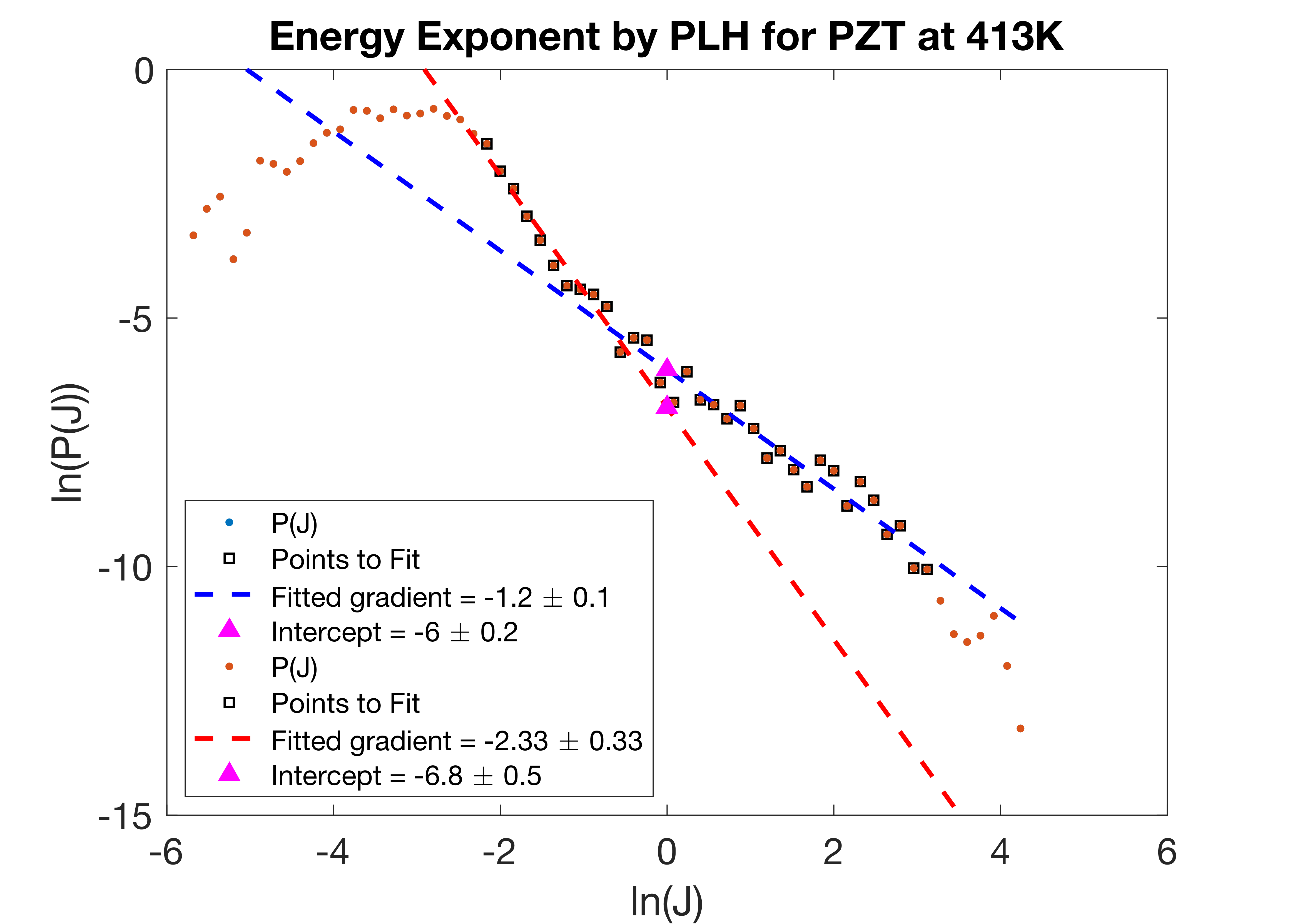}%
}
\subfloat[Energy Maximum-likelihood PZT at 413 K \label{fig:413_ml}]{%
  \includegraphics[width=8.6cm]{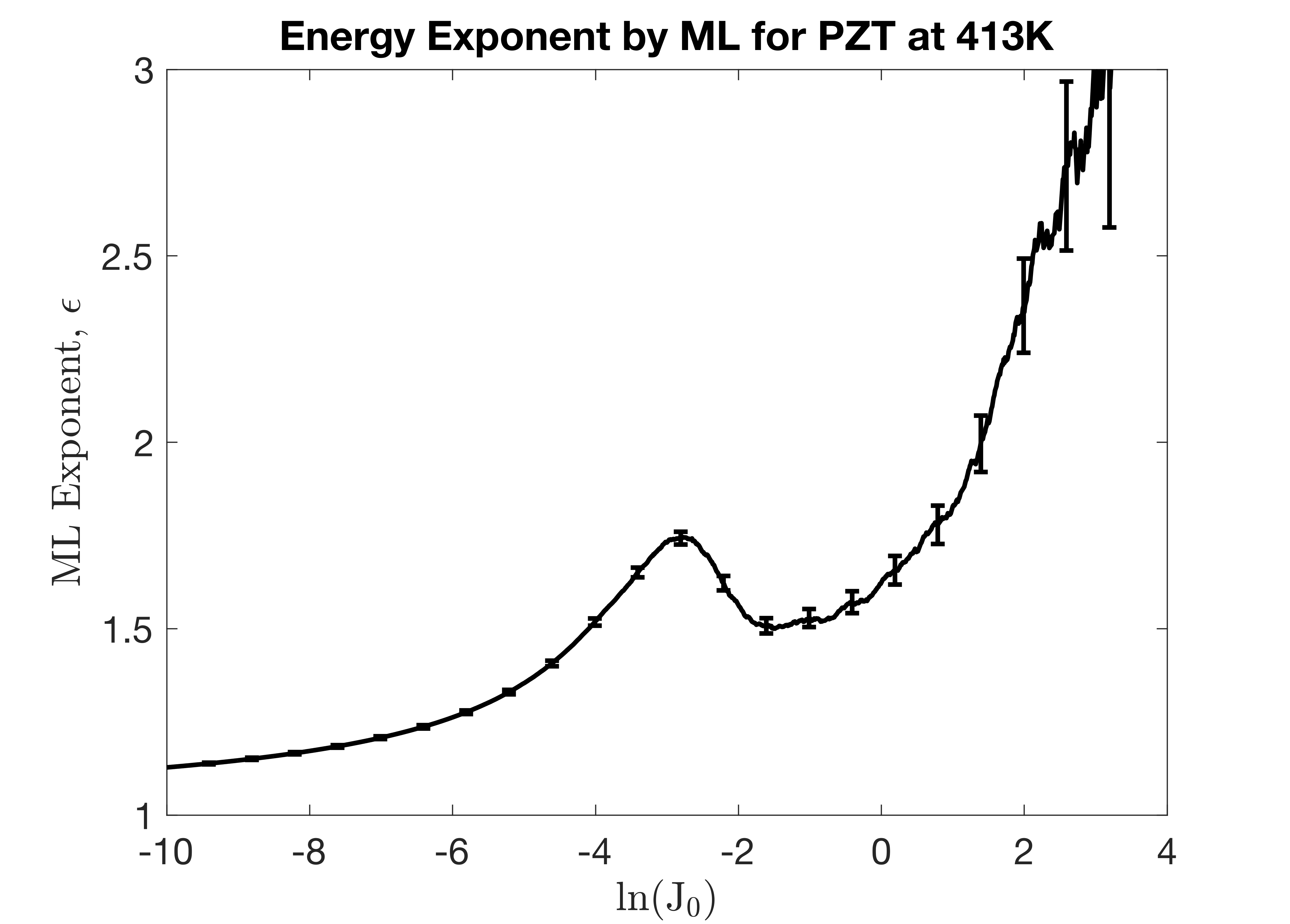}%
}
\caption{Illustrative examples of PLH and ML plots for behaviour groups A and B. \textbf{(a)} shows PLH and \textbf{(b)} shows ML plot for energy exponent of PZT sample at 259 K (group A). \textbf{(c)} shows PLH and \textbf{(d)} shows ML plot for energy exponent of PZT sample at 413 K (group B). Characteristics of group A are the flat nature of the ML plot and the two exponents with a small angle between them in the PLH. Characteristics for group B are single large peak in the ML plot before the plateau and two exponents with large angle between them on the PLH.}
\label{fig:energy_vt}
\end{figure}

\twocolumngrid

\begin{table}[h]
    \centering
    \begin{tabular}{|c|c|c|c|}
    \hline
        Temperature, K & $\epsilon_1$ & $\epsilon_2$ & $\epsilon_3$ \\
         \hline
        160 & 1.29 $\pm$ 0.10 & 2.10 $\pm$ 0.21 & \\
        227 & 1.26 $\pm$ 0.21 & 1.68 $\pm $0.18 & \\
        259 & 1.30 $\pm$ 0.06 & 1.80 $\pm$ 0.12 & \\
        323 & 1.27 $\pm$ 0.17 & 2.19 $\pm$0.17 & 2.23 $\pm$ 0.61 \\
        333 & 1.67 $\pm$ 0.13 & 1.92 $\pm$ 0.19 & \\
        353 & 1.54 $\pm$ 0.11 & 1.92 $\pm$ 0.35 & \\         
        373 & 0.99 $\pm$ 0.36 & 2.01 $\pm$ 0.54 & 2.32 $\pm$ 0.43 \\
        393 & 1.24 $\pm$ 0.08 & 2.64 $\pm$ 0.25 & \\
        413 & 1.20 $\pm$ 0.10 & 2.33 $\pm$ 0.33 & \\
        423 & 1.48 $\pm$ 0.14 & 2.06 $\pm$ 0.64 & 2.52 $\pm$ 0.21 \\
        433 & 1.03 $\pm$ 0.18 & 2.09 $\pm$ 0.21 & 2.30 $\pm$ 0.24 \\
        453 & 1.31 $\pm$ 0.25 & 1.81 $\pm$ 0.09 & \\
        473 & 1.48 $\pm$ 0.14 & 1.9 $\pm$ 0.13 & \\
        \hline
    \end{tabular}
    \caption{Summary of energy exponents obtained from power-law histograms at various temperatures.}
    \label{tab:energy_vt}
\end{table}

\section{Energy Exponents for BaTiO$_{3}$ in tetragonal and rhombohedral phases}

\begin{figure}[h!]
\subfloat[\centering Power-law Histogram pBTO at RT]{%
  \includegraphics[width=8.6cm]{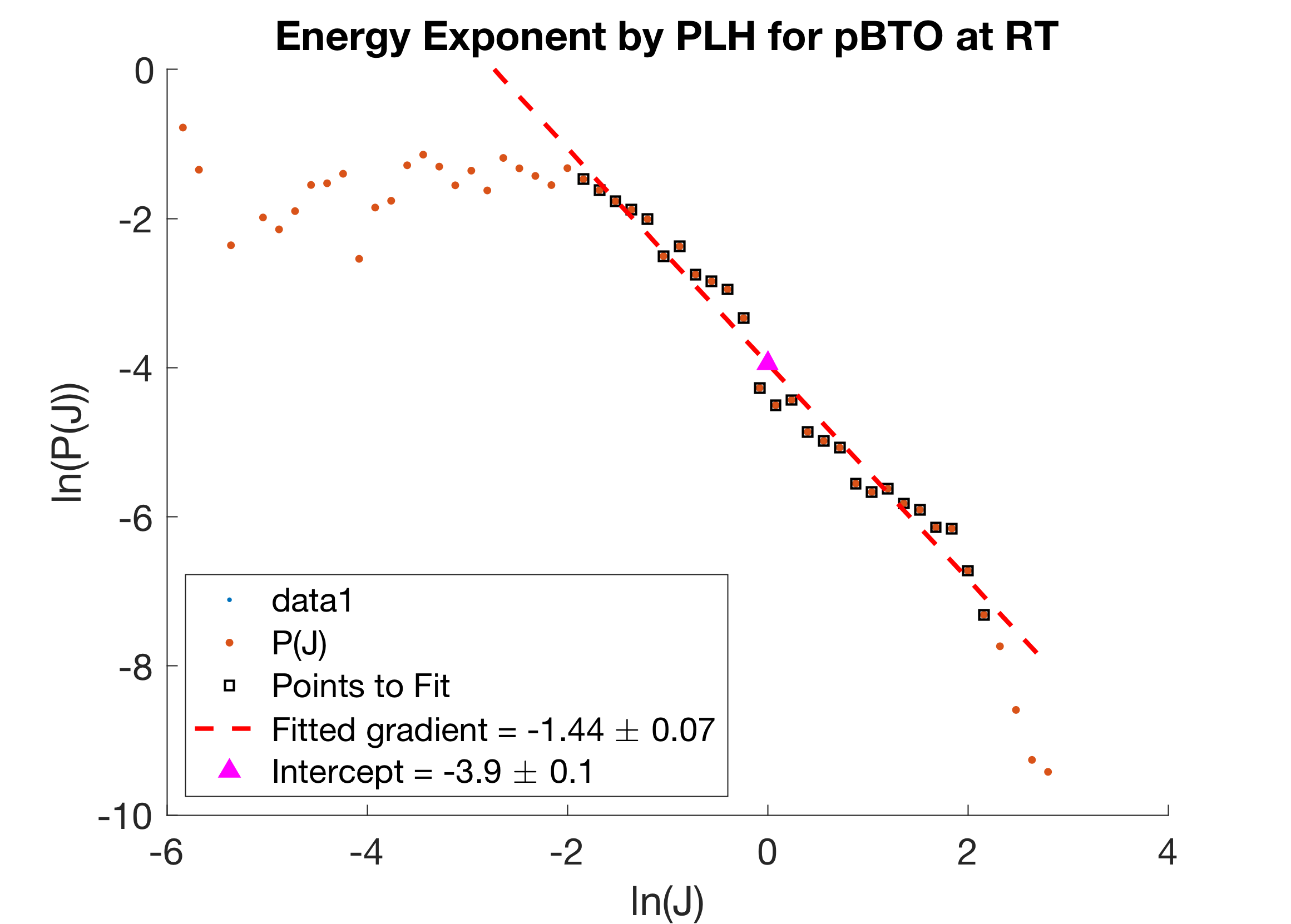}%
}

\subfloat[Maximum-likelihood pBTO at RT\label{fig:bto_ml_rt}]{%
  \includegraphics[width=8.6cm]{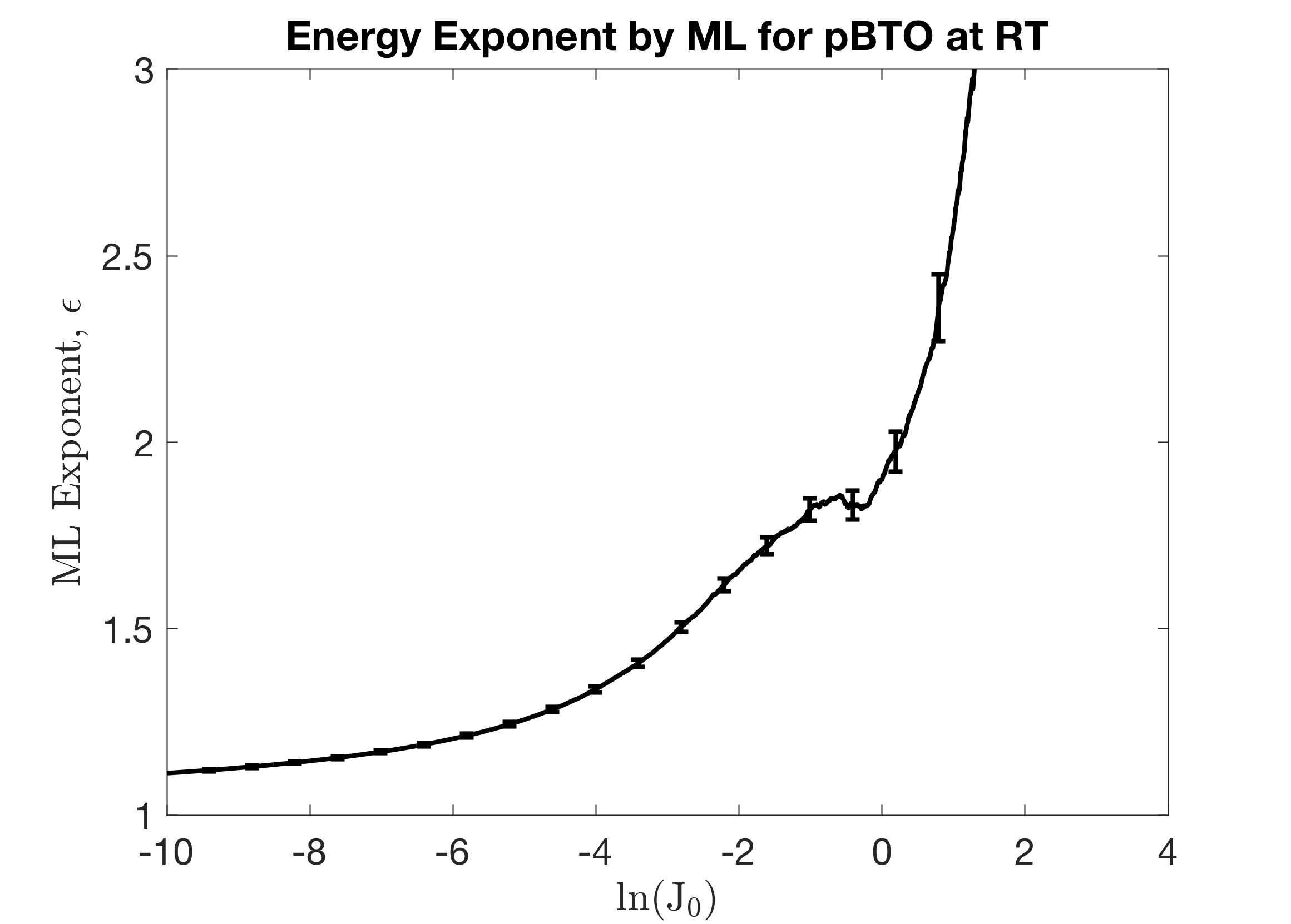}%
}
\caption{\textbf{(a)} Power-law histogram and \textbf{(b)} maximum-likelihood plot for energy exponent of pBTO sample at room temperature. PLH seems to show a power-law with $\epsilon=1.4\pm0.1$ but there is no plateau on ML curve.}
\label{fig:bto_rt}
\end{figure}

\begin{figure}[h!]
\subfloat[\centering Power-law Histogram pBTO at 80 K]{%
  \includegraphics[width=8.6cm]{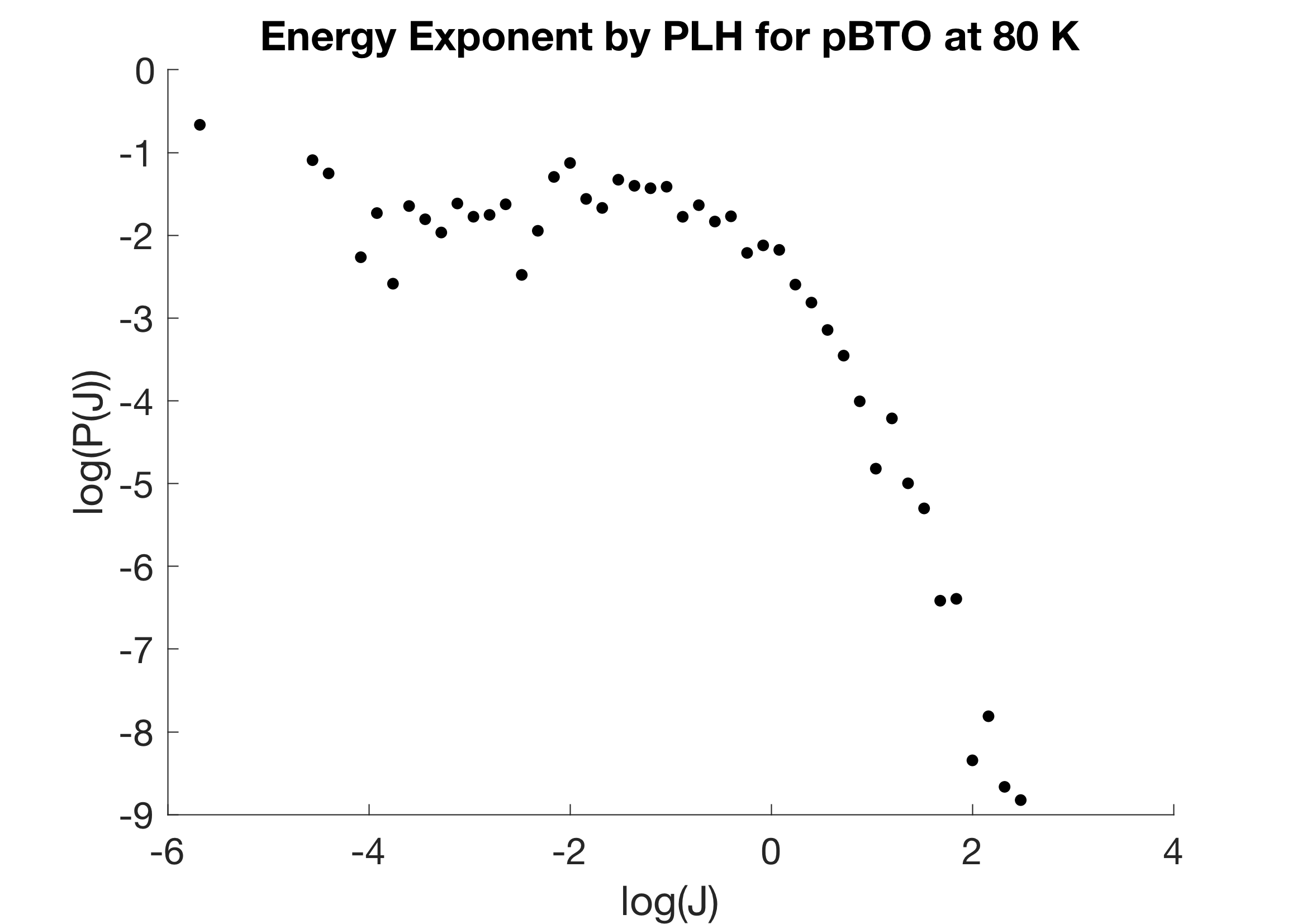}%
}

\subfloat[Maximum-likelihood pBTO at 80 K\label{fig:bto_ml_80}]{%
  \includegraphics[width=8.6cm]{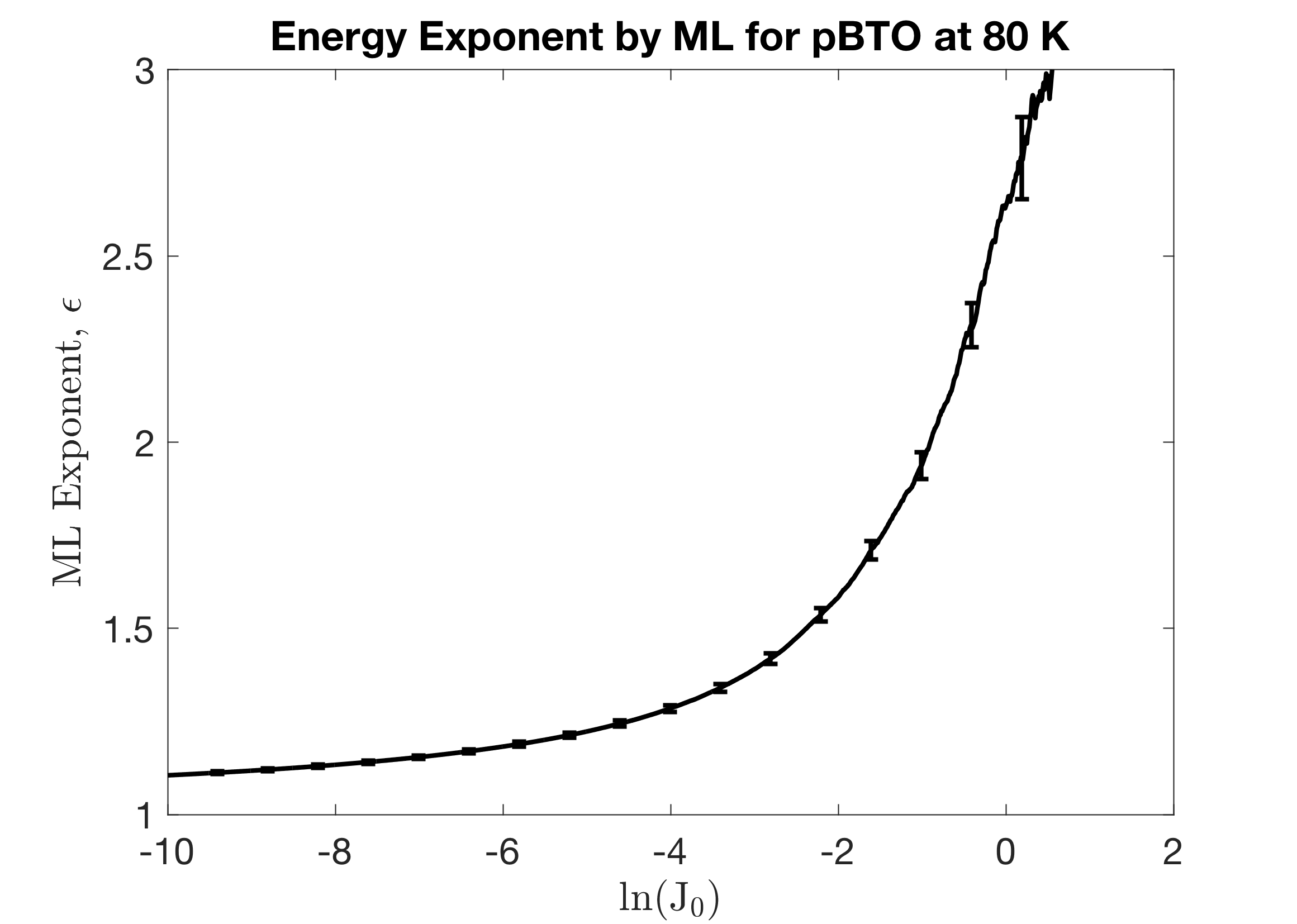}%
}
\caption{\textbf{(a)} Power-law histogram and \textbf{(b)} maximum-likelihood plot for pBTO sample at 80 K. There is no straight line in the PLH and no plateau in the ML curve, hence this distribution does not obey a power-law.}
\label{fig:bto_80}
\end{figure}

Fig. \ref{fig:bto_rt} illustrates our aftershock data in tetragonal single-crystal barium titanate at room temperature. And Fig. \ref{fig:bto_80} shows data at T=80K.  The latter do not exhibit a linear dependence in the log-log graph of pulse height; and in comparison with ambient data, the plateau is much larger and the fall-off more abrupt yet rounded and nonlinear.  Neither is there is plateau in the maximum likelihood graph.  We conclude from this that the Barkhausen switching dynamics are very different at T=80 K and might not not be characterized by a single exponential dependence.  If there is a power-law dependence, it sets in at only high slew rates J. This may be related to the recent observations \cite{Mohamad1982AcousticGermanate} in lead germanate.  Other examples of ferroelectrics with very high coercive field thresholds for Barkhausen noise include \cite{Kumari2016Palladium-basedExperiment} GaFeO3.   See also the results by Y. Tokura’s group \cite{Kagawa2016QuantumComplex} that domain wall creep thermally vanishes in ferroelectrics at low temperatures  (replaced by tunneling at the lowest T).

Lowering the temperature to T=80K into the rhombohedral phase gave very different Barkhausen data (Fig. \ref{fig:bto_80}).  Doping barium titanate with Fe \cite{QIU2010PhaseCeramics,Xu2009Room-temperature3} gave too soft a material for Barkhausen data.

\section{Summary}

The present work has resulted in accurate values for the Omori exponent of aftershocks in Barkhausen pulses in PZT, not previously reported.  The value obtained is very close to unity, as predicted, and more accurate than previously reported values.

Rather more surprising are the strong temperature dependences for the energy exponents in both PZT and barium titanate; these show as temperature is changed an evolution to two distinct superimposed values, one near 2.0 and one near 1.4.  In barium titanate no maximum likelihood exponent can be fitted in its rhombohedral R3m phase at low temperatures (T=80K).

Our suggestion is that in general at least two depinning processes are involved, one involving point defects and the other, extended defects (threading dislocations in PZT). Another possibility [z’] is that one set is Barkhausen noise due to domain walls and the other set due to thermal microcracking.

\begin{enumerate}
    \item \textbf{Omori data on PZT:}\\ p = 0.95$\pm$0.03 (large pulse, ca. 500-1000 attoJ); p = 0.94$\pm$0.06 (small pulses, ca. 1-10 attoJ)
    \item \textbf{Temperature data on PZT:}
            \begin{enumerate}[a)]
                \item Room temperature (three values for energy exponent): 1.79$\pm$0.08 (lowest slew rate), 1.47$\pm$0.04 (intermediate; this may be a transition regime rather than a separate exponent), 2.04$\pm$0.19 (highest slew rate)
                \item Energy exponents at T=160K:  2.10$\pm$0.20 and 1.69$\pm$0.10 
                \item Energy exponents at T=227K:  1.26$\pm$0.21 and 1.68$\pm$0.18 
                \item Amplitude exponent: 2.23$\pm$0.11
                \item Duration exponent (no unambiguous power law; insufficient data)
            \end{enumerate}
    \item \textbf{Temperature data on BaTiO$_{3}$:}\\ Aftershocks display very different statistics in rhombohedral barium titanate  at T=80K, with a long plateau of steady-state followed by abrupt cessation -- not a simple power-law dependence according to maximum likelihood analysis. \cite{Prabakar2004PhaseEmission}  A similar effect was reported some years ago in lead germanate but is not understood.  It suggests that the domain motion kinetics in high-coercive field ferroelectrics are limited by nucleation times and not by field-activated creep.
    
\end{enumerate}

\begin{acknowledgments}
We thank Ekhard Salje, Finlay Morrison, and Jonathan Gardner for helpful discussions.  Work supported by EPSRC grant EP/PO24637/01.
\end{acknowledgments}

\bibliography{references1}

\end{document}